%% file: KrzysztofPhD.tex
\documentclass[twoside,reqno]{article}
\usepackage{amsmath}
\usepackage{amsthm}
\usepackage{enumerate}
\usepackage{amssymb}
\usepackage{amsmath}
\usepackage{amsfonts}
\usepackage{bm} 

\usepackage{verbatim}
\usepackage{url}
\usepackage[latin1]{inputenc}

\usepackage{caption}
\usepackage{graphicx,color}
\usepackage{pdflscape}
\usepackage{float}
\usepackage[final]{pdfpages}

\usepackage[hmarginratio=1:1]{geometry}
\usepackage{multirow}

\usepackage[round]{natbib}
\usepackage[nottoc,numbib]{tocbibind}

\usepackage{xr}
\usepackage{algorithmic}
\usepackage{algorithm}
\usepackage{listings}

\newcommand{\ud}{\mathrm{d}}
\newcommand{\bit}{\begin{itemize}}
\newcommand{\eit}{\end{itemize}}

\newcommand\independent{\protect\mathpalette{\protect\independenT}{\perp}}
\def\independenT#1#2{\mathrel{\rlap{$#1#2$}\mkern2mu{#1#2}}}
\newcommand{\be}{\begin{equation}} \newcommand{\ee}{\end{equation}}
\newcommand{\bd}{\begin{displaymath}} \newcommand{\ed}{\end{displaymath}}
\newcommand{\E}[1]{\operatorname{E}\left[ #1 \right]}

\newcommand{\var}[1]{\operatorname{Var}\left[ #1 \right]}
\newcommand{\cov}[2]{\operatorname{Cov}\left[ #1,#2 \right]}

\newcommand{\Cov}[1]{\operatorname{Var}\left[ #1 \right]}

\newcommand{\prob}[2]{\operatorname{P}\left( #1 = #2 \right)}

\def\be{\begin{equation}}
\def\ee{\end{equation}}
\def\bea{\begin{equation*}}
\def\eea{\end{equation*}}

\newtheorem{theo}{Theorem}[section]

\theoremstyle{definition}

\theoremstyle{remark}
\newtheorem{preremark}[theo]{Remark}
\newtheorem{preex}[theo]{Example}

\numberwithin{equation}{section}

\date{\today}
\author{Krzysztof Bartoszek}

\def\authorname{Krzysztof Bartoszek}

\def\longtitle{Stochastic models in phylogenetic comparative methods: analytical properties and parameter estimation}
\def\longtitleSV{Stokastiska modeller f\"or fylogenetiska komparativa metoder -- analytiska egenskaper och parameteruppskattning}

\def\keywords{\noindent{\bf Keywords:}
Adaptation, Allometry, Birth--death process, Branching diffusion,
Brownian motion, Conditioned branching process, Evolution, General Linear Model, 
Hybridization, Macroevolution, 
Measurement error, Multivariate phylogenetic comparative method, Optimality, Ornstein--Uhlenbeck process,  
Phyletic gradualism,
Phylogenetic inertia, Phylogenetic uncertainty, Punctuated equilibrium, 
Yule tree 
}
\def\keywordsSV{\noindent{\bf Nyckelord:}
Allometri, Anpassning, Brownsk r\"orelse, Evolution, 
Fylogenetisk os\"akerhet, Fylogenetisk tr\"oghet,
F\"odelse--d\"odsprocess, F\"orgrenad diffusion, General Linear Model, Hybridisering,
Betingad f\"orgrenings\-process, 
Makroevolution,
Multivariat fylogenetisk komparativ metod, M\"at\-fel,
Optimalitet, Ornstein--Uhlenbeckprocess,
Phyletic gradualism,  Punctuated equilibrium, 
Yuletr\"ad
}

\def\degree{Doctor of Philosophy}
\def\division{Division of Mathematical Statistics}
\def\department{Department of Mathematical Sciences}
\def\cth{Chalmers University of Technology\\and University of Gothenburg}
\def\gbg{G\"oteborg, Sweden 2013}
\def\post{SE-412 96 G\"OTEBORG, Sweden}
\def\phone{Phone: +46 (0)31-772 10 00}
\def\authorsmail{bartoszekkj@gmail.com}
\def\printservice{Typeset with \LaTeX.\\\department\\Printed in G\"oteborg, Sweden 2013}

\begin{document}

\pagenumbering{roman}
\include{titlepage}

\include{abstract}
\cleardoublepage

\include{abstractSV}
\cleardoublepage

\include{acknow}
\cleardoublepage

\include{papers}
\cleardoublepage

\include{papersExtra}
\cleardoublepage

\tableofcontents
\cleardoublepage

\pagenumbering{arabic}
\setcounter{page}{1}

\include{phd}

\newpage
\bibliography{compmethod,mvslouch,compbio,sde,appendixMError,lic,phdProjBishopsWarblers,matrix,treefree,phylnet}
\bibliographystyle{plainnat}




\vspace{1cm}
\end{document}

%% file: titlepage.tex
{\setlength{\parindent}{0pt}
\thispagestyle{empty}

\begin{center}
{\sc Thesis for the Degree of \degree}

\vspace{4cm}

\parbox{\textwidth}{\center\LARGE\bf\longtitle}

\vspace{1.5cm}

{\Large\sc\authorname}

\end{center}

\vfill
\begin{center}
\begin{figure}[htbp]
\includegraphics[width=0.9\textwidth]{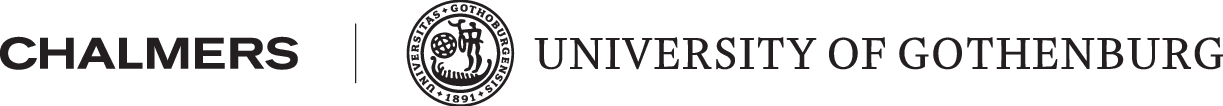}
\end{figure}
{\it\division}\\
{\it\department}\\
{\sc\cth}\\
\gbg
\end{center}

\pagebreak

\thispagestyle{empty}

{\bf{\small\longtitle}}\\
{\it \authorname}\\

\vspace{3mm}

Copyright \copyright\ \authorname, 2013.

\vspace{5mm}

ISBN 978--91--628--8740--7\\

\vspace{3mm}

\vfill

\department \\
\division\\
\cth\\
\post\\
\phone

\vspace{\baselineskip}

Author e-mail: {\tt \authorsmail}

\vspace{\baselineskip}

\printservice
}

%% file: abstract.tex
\noindent
{\bf{\small\longtitle}}\\
{\it \authorname}\\

\begin{center}
{\bf Abstract}
\end{center}





Phylogenetic comparative methods are well established tools for using
inter--species variation to analyse phenotypic evolution and adaptation.
They are generally hampered, however, by predominantly univariate approaches and
failure to include uncertainty and measurement error in the phylogeny as well as
the measured traits. This thesis addresses all these three issues. 

First, by
investigating the effects of correlated measurement errors on a phylogenetic regression.
Second, by developing a multivariate Ornstein--Uhlenbeck model
combined with a maximum--likelihood estimation package in R. This model allows,
uniquely, a direct way of testing adaptive coevolution. 

Third, accounting for the often substantial
phylogenetic uncertainty in comparative studies requires an explicit model for the tree.
Based on recently developed conditioned branching processes, with Brownian and Ornstein--Uhlenbeck
evolution on top, expected species similarities are derived, together with phylogenetic confidence
intervals for the optimal trait value. Finally, inspired by these developments, the phylogenetic framework is
illustrated by an exploration of questions concerning ``time since hybridization'', the distribution of which
proves to be asymptotically exponential.

\vfill
\keywords

%% file: abstractSV.tex
\noindent
{\bf{\small\longtitleSV}}\\
{\it \authorname}\\

\begin{center}
{\bf Sammanfattning}
\end{center}





Fylogenetiska j\"amf\"orande metoder \"ar v\"al etablerade verktyg f\"or analys
av fenotypisk evolution och adaptation baserat p\aa\ variation mellan arter.
Generellt \"ar emellertid dessa metoder begr\"ansade av \"overv\"agande univariata
modeller och avsaknad av m\"ojligheter att inkludera os\"akerheter och m\"atfel i s\aa v\"al
fylogeni som i de analyserade egenskaperna. I den h\"ar avhandlingen behandlas
alla dessa aspekter. F\"orst genom att studera effekterna av korrelerade m\"atfel p\aa\ 
en fylogenetisk regression--analys. D\"arefter genom att utveckla en multivariat
Ornstein--Uhlenbeck--modell f\"or adaptiv evolution, med parameter--skattningar i R.
Denna modell till\aa ter, unikt och f\"or f\"orsta g\aa ngen, direkt analys av adaptiv
samevolution mellan egenskaper. 

Att ta h\"ansyn till den ofta avsev\"arda fylogenetiska
os\"akerheten i j\"amf\"orande studier kr\"aver en explicit modell f\"or det fylogenetiska tr\"adet.
H\"ar anv\"ands ist\"allet de nyutvecklade ``betingade f\"orgreningsprocesserna'' 
i kombination med ``Brownian motion'' och ``Ornstein--Uhlenbeck''
processer s. a. s. ``ovanp\aa ''. Ur detta h\"arleds t. ex. ``hur lika'' arter kan f\"orv\"antas vara, samt
en sorts fylogenetiska konfidensintervall kring det optimala egenskapsv\"ardet.

Slutligen, inspirerad av ovanst\aa ende metodutveckling, ges en till\-\"amp\-ning
p\aa\  aktuella fr\aa gor kring ``tid sedan hybridisering'', vars f\"ordelning visar sig vara
asymptotiskt exponentiell.

\vfill
\keywordsSV

%% file: acknow.tex
\begin{center}
{\bf Acknowledgments}
\end{center}
I would like to thank my parents and Ewa \L \c aczy\'nska for their constant support during the 
time of my PhD studies.
~\\~\\
\noindent
I would like to thank Staffan Andersson, Petter Mostad, Olle Nerman, Igor Rychlik and Serik Sagitov for their 
immense help 
in writing this thesis and Patrik Albin, and Thomas Ericsson for many helpful comments.
~\\~\\
\noindent
I would also like to thank Staffan Andersson, Mohammad Asadzadeh, Jakob Augustin, Wojciech Bartoszek, 
Joachim Domsta, 
Peter Gennemark, Thomas F. Hansen, Graham Jones, Micha\l\ Krze\-mi\'n\-ski, Petter Mostad, Bengt Oxelman, Pietro Li\'o, 
Jason Pienaar, 
Krzysztof Podg\'orski, Maria Prager, 
Ma\l gorzata Pu\l ka, Serik Sagitov, Jaros\l aw Sko\-kow\-ski, Anil Sorathiya, Franciska Steinhoff, Anna Stokowska and 
Kjetil Voje for fruitful collaborations.
~\\~\\
\noindent
I finally thank my colleagues at the Department of Mathematical Sciences for a wonderful
working environment.

\begin{flushright}
\authorname\\
G\"oteborg, \today
\end{flushright}

%% file: papers.tex
\begin{center}
{\bf List of Papers}
\end{center}
The PhD thesis is based on the following papers,
\begin{enumerate}[I.]
\item Hansen T. F. and \textbf{Bartoszek K.} (2012).
Interpreting the evolutionary regression: the interplay between observational
and biological errors in phylogenetic comparative studies. 
Systematic Biology, \textbf{61}(3):413-425.
\item \textbf{Bartoszek K.}, Pienaar J., Mostad P., Andersson S. and Hansen T. F. (2012).
A phylogenetic comparative method for studying multivariate adaptation.
Journal of Theoretical Biology, \textbf{314}:204-215.
\item Sagitov S. and \textbf{Bartoszek K.} (2012).
Interspecies correlation for neutrally evolving traits.
Journal of Theoretical Biology, \textbf{309}:11-19.
\item \textbf{Bartoszek K.} and Sagitov S. (\textit{submitted}).
Phylogenetic confidence intervals for the optimal trait value. 
\item \textbf{Bartoszek K.} (\textit{submitted}).
Quantifying the effects of anagenetic and cladogenetic evolution.
\item \textbf{Bartoszek K.}, Jones G., Oxelman B. and Sagitov S. (2012).
Time to a single hybridization event in a group of species with unknown ancestral history.
Journal of Theoretical Biology, \textbf{322}:1-6.
\end{enumerate}
                                                                                                                                                

%% file: papersExtra.tex
\begin{center}
{\bf List of papers not included in this thesis}
\end{center}
\begin{enumerate}[I.]
\setcounter{enumi}{7}
\item \textbf{Bartoszek K.} and Stokowska A. (2010).
Performance of pseudo--likelihood estimator in modelling cells' proliferation with noisy measurements.
Proceedings of the XII International Workshop for Young Mathematicians ``Probability Theory and Statistics'', Cracow: pp. 21-42.
\item \textbf{Bartoszek K.}, Krzemi\'nski M., and Skokowski J. (2010).
Survival time prognosis under a Markov model of cancer development.
Proceedings of the Sixteenth National Conference on Applications of Mathematics in Biology and Medicine, Krynica: pp. 6-11.
\item  Asadzadeh M. and \textbf{Bartoszek K.} (2011).
A combined discontinuous Galerkin and finite volume scheme for multi--dimensional VPFP system.
AIP Conference Proceedings, \textbf{1333}:57-62.
\item Steinhoff F. S., Graeve M., \textbf{Bartoszek K.}, Bischof K. and Wiencke C. (2012).
Phlorotannin Production and Lipid Oxidation as a Potential Protective Function Against High Photosynthetically Active and UV Radiation in Gametophytes of \textit{Alaria esculenta} (Alariales, Phaeophyceae).
Photochemistry and Photobiology, \textbf{88}(1):46-57.
\item \textbf{Bartoszek K.} (2012).
The Laplace motion in phylogenetic comparative methods.
Proceedings of the Eighteenth National Conference on Applications of Mathematics in Biology and Medicine, Krynica Morska: pp. 25-30.
\item \textbf{Bartoszek K.} and Krzemi\'nski M. (\textit{in press}).
Critical case stochastic phylogenetic tree model via the Laplace transform. Demonstratio Mathematica.
\end{enumerate}
                                                                                                                                                

%% file: phd.tex
\section{Introduction}\label{secIntro}
\subsection{Phylogenetic comparative methods}
In evolutionary biology one of the fundamental concerns  is how different inherited traits  evolve, depend  on each
other and react to the changing environment. The natural approach to answer questions related to these problems
is to record trait measurements and environmental conditions, and subsequently analyze them.

There is however a problem 
that is not taken into account with just this approach. Namely, the sample is not independent as the different
species are related by a common evolutionary history. 
The immediate consequence of this common evolutionary history is 
that closely related species will have similar trait values. As it is very probable
that closely related species will be living in similar environments 
we could
observe a false dependency between traits and environment. One needs to take into account this common
phylogenetic history to be able to distinguish between effects stemming from similarity
and evolutionary effects. 
This is of particular importance when studying the adaptation of traits towards
environmental niches. Are species living in a similar environment similar because
they have adapted to it or is it because they are closely related?

In statistical terms forgetting about the inter--species dependency
means that we would be analyzing data  under a wrong model.
Such an analysis from the perspective of the correct model 
could lead to biased estimates and misleading parameter interpretations.
It would certainly result in wrong confidence intervals and p--values. 

The thesis presented here considers only a very specific part of the extremely wide field of phylogenetics
--- the mathematics and computations involved in modelling phenotype evolution 
and estimation of parameters of evolutionary models for continuous comparative data. 
A general overview of the field of phylogenetics has been presented by e.g. \citet[][]{JFel2004,PLemMSalAVan2009}.
Biological background for phylogenetic comparative methods has been discussed by e.g.
\citet{PHarMPag} with methodologies for discrete data and simple continuous settings.
Evolution of discrete characters is also discussed by \citet{MPag1994,MPag1999}
and \citet{EPar} discusses a versatile R \citep{R} package (ape) with
many example analysis.


This thesis is based on six papers. 
The first two, concerning parameter estimation, each have 
R code attached to them.

Paper I \cite{THanKBar2012} concerns the problem of measurement error 
(or observational variance, as most comparative studies take average values from a number
of individuals) in phylogenetic regression studies. The problem of measurement error
is widely addressed in the literature \citep[see e.g.][]{WFul1987,LGle1992,JBuo}.
However the general case of dependent
errors in the predictor variables or dependently evolving predictor variables 
seems to be lacking in literature. In Paper I we consider general covariance
matrices $\mathbf{V}_{d}$ and $\mathbf{V}_{u}$ relating the predictors and measurement
errors between observations. 
We show that if one assumes some
structure on these matrices then substantial simplifications can be made in the bias
formula. 
Correcting for the bias can increase the mean square error of the estimate
so we  propose a criterion,
depending on the observed data, that indicates whether correction is
beneficial or not. 

Paper II \citep{KBarJPiePMosSAndTHan2012} presents and develops a multivariate model of trait
evolution based on an Ornstein--Uhlenbeck (OU) 
\citep[][]{JFel1988,THan1997,MButAKinOUCH,THanJPieSOrzSLOUCH}
type stochastic process,
often used for studying trait adaptation, co--evolution, allometry or
trade--offs.
Included with the paper is an R software package
to estimate the model's parameters. 
We are not limited to studying interactions between
traits and environment. Within the presented multivariate framework it
is possible to pose and rigorously test
hypotheses about adaptation or trade--offs
taking into account the phylogeny.

In the classical phylogenetic comparative methods setting it is assumed that 
the phylogeny is fully resolved
\citep[however this has not always been the case][]{AEdw1970}. 
This might seem reasonable as 
with the current wealth of molecular information we 
can obtain more and more accurate trees. However there is 
interest in tree--free methods \citep{FBok2008}.
In many cases the tree itself might not be of direct interest, in fact it could actually be 
in some situations a nuisance parameter, motivating 
methods that preserve distributional properties of the observed phenotypic value
without a dependence on a particular tree \citep{FBok2010}. 
We still have unresolved clades, 
for example in the Carnivora order
\citep[used for an example analysis by][]{FCraMSuc2013}.
Another example for the usefulness of tree--free methods is in the analysis
of fossil data \citep[][underline the importance of incorporating fossil data in comparative studies]{GSlaLHarMAlf2012}. 
There may be available rich fossilized phenotypic information but the 
molecular material might have degraded so much
that it is impossible to infer evolutionary relationships.

\citet{FCraMSuc2013} in their discussion suggest that with the advancement of our 
knowledge of the tree of life interest in tree--free methods may possibly diminish along
with the number of unresolved clades. 
Firstly, estimation of parameters of complex (even as ``simple'' as the Ornstein--Uhlenbeck one) 
evolutionary models can take intolerable 
a\-mounts of time for large phylogenies. Therefore unless there will be 
a major speed--up in estimation algorithms and computing power (but this will also probably
come at a financial and energetic cost) tree--free methods should be 
considered as a viable alternative. 
Secondly, in situations where we can do a tree--based analysis
``integrating'' over the phylogeny can be used as a sanity check,
whether the conclusions based on the inferred phylogeny deviate much from those
from a ``typical'' phylogeny. Tree--based methods very rarely
use analytical estimators, they are
currently nearly
always either a numerical optimization of the likelihood
function or MCMC/simulation based (Approximate Bayesian Computations).
Such estimation algorithms, especially in a high--dimension parameter space
are always at risk of falling into a local minimum. 
Therefore thirdly, contradictory tree--free results will
give a reason for investigation.
Additionally numerical and Bayesian methods nearly always require
a seed/prior distribution to start the estimation. Calculating these using tree--free
methods might be an attractive and fast option for the researcher. 

Tree--free methods naturally require a modelling framework for
the phylogeny. The one of conditioned (on the number of extant tips) branching processes is an appealing option.
We know how many species we have observed. This statement of course raises another statistical question 
that still needs to be addressed --- what if this number is only a random draw from the true number of contemporary species,
this can be the case especially with lower orders i.e. below simple invertebrates or primitive plants.
Conditioned branching processes, motivated amongst other by phylogenetic questions,
have received significant attention in the past decade. In particular the research
of \citet{MSteAMcK2001,LPop2004,DAldLPop2005,TGer2008a,TSta2008,TSta2009,TSta2011b,KHarDWonTSta2010,ALam2010,AMooOGasTStaHLiMSte2012}
has provided tools for us to develop tree--free phylogenetic comparative methods. 

Despite the long time that phylogenetic comparative methods have been around 
not many analytical statistical properties of them have been found.
\citet{CAne,LHoCAne2013} could be the most notable exceptions where 
the consistency of phylogenetic maximum likelihood estimators is studied. 
However they differ from our work by a different model of tree growth and
treatment of the ancestral state. This multitude of possible modelling 
assumptions all having their own biological justification makes 
phylogenetic comparative methods exciting to study and we hope that our conditioned branching process approach
will aid in understanding them. 

In line with the current comparative methods framework we consider 
three evolutionary models on top of our unobserved phylogeny. In Paper III \citep{SSagKBar2012}
we consider the Brownian motion model \citep{JFel1985}.
This was the first continuous comparative model introduced  but
\citet{AEdw1970} had already considered the likelihood for a Brownian motion on top of an unobserved pure birth--tree.
For it we calculate the interspecies correlation coefficient --- how similar do we 
expect two randomly sampled species to be. Paper IV moves on to
the Ornstein--Uhlenbeck model \citep{JFel1988,THan1997,MButAKinOUCH,THanJPieSOrzSLOUCH,KBarJPiePMosSAndTHan2012} 
interpreted by us as an adaptive model. We assume a single--peak model, calculate 
for the pure--birth tree phylogenetically
corrected confidence intervals for the optimal trait value
and derive central limit theorems for the sample mean. Interestingly
the form of the correction and central limit theorem depends
on the value of $\alpha$ (the adaptation rate) with a phase 
transition occurring at $\alpha$ equalling half the speciation rate.

In Paper V we add to the Brownian motion and Ornstein--Uhlenbeck evolutionary models
a jump component at speciation events \citep[][]{AMooDSch1998,AMooSVamDSch1999,FBok2002,FBok2008,FBok2010}.
Our study indicates that speciation jump events decorrelate species and we also postulate
that using quadratic variation is convenient to compare effects of gradual and punctuated evolution.

To derive analytics of tree--free models one has to study probabilistic properties
of coalescent times of phylogenetic trees generated by birth--death models. 
These results, some of which, like the Laplace transform of the height of a pure birth tree
(Paper IV) we believe are novel themselves, open the way to study the statistical
properties of various tree indices (e.g. the total cophenetic index) and 
metrics on the space of trees. Such properties are needed 
so one can rigorously compare trees and ask if they fit a particular
speciation--extinction model \citep{OPybPHar2000,JFel2004}. These
have been of course studied previously in particular by
\citet{GCarMLlaFRosGVal2010,WMul2011,GCarAMirFRos2012,AMirFRosLRot2013}. 
Their approach however has been heavily influenced by number theory 
resulting in very general (in terms of models) but lengthy proofs.
Our approach on the other hand usually results in 
derivations which are very short  but dependent 
on knowing the properties of coalescent times under a particular branching process
(see Papers III, V).

The final paper underlying this thesis, Paper VI \citep{KBarGJonBOxeSSag2013}, is in a different spirit but has been motivated
by our results concerning conditioned branching processes. We model a single hybridization
event inside a pure--birth tree. We assume that for each fixed $n$ there is 
a constant hybridization rate, $\beta_{n}$ between each pair of species. Then
if $n\beta_{n}/\lambda \to 0$, where $\lambda$ is the speciation rate, 
we obtain that the distribution of the time till the hybridization event
converges to the exponential distribution with rate $2\lambda$.


\subsection{Stochastic differential equations}
There are two components of our tree--free framework. The first one 
is the model for the randomly evolving phenotype. The two currently standard models
(see Section \ref{sbsecPapII} for their definitions)
are the Brownian motion model \citep{JFel1985} and the Ornstein--Uhlenbeck model \citep{THan1997}.
These two stochastic processes appear as solutions of stochastic differential equations (SDEs) given in the form,
\be\label{eqsde}
\ud X(t) = \mu(t,X(t))\ud t + \sigma(t,X(t)) \ud W(t),
\ee
with initial condition $X_{0}$. $W(t)$ is the standard Wiener process
--- independent increments with \mbox{ $W(t)-W(s) \sim
\mathcal{N}(0,t-s), 0 \le s \le t$}. Traditionally Eq. \eqref{eqsde} is interpreted as,
\be\label{eqsdeIto}
X(t) = X(0) + \int\limits_{0}^{t} \mu(u,X(u))\ud u + \int\limits_{0}^{t} \sigma(u,X(u)) \ud W(u),
\ee
where we understand $\int_{0}^{t} Y(u) \ud W(u)$ as the It\^o stochastic integral. 
We assume that $t \in [0,T]$ for some (large) value of $T$.
The sample paths of the  process $Y(u)$ (notice as $\sigma(u,X(u))$ depends on $X$ it is a stochastic process)
have to satisfy (apart from the measurability and adaptedness of $Y$),
\bd
P\left(\int\limits_{0}^{T} Y^{2}(u) \ud  u< \infty \right) = 1
\ed
and the It\^o integral is the limit in probability,
\bd
\int_{0}^{t} Y_{n}(u) \ud W(u) \stackrel{P}{\longrightarrow} \int_{0}^{t} Y(u) \ud W(u),
\ed
for a sequence of simple (piecewise constant, adapted and measurable) processes 
(on a predefined partition sequence 
\bd
0=t_{0}<t^{n}_{1}<\ldots<t^{n}_{k-1}<t^{n}_{k}<t^{n}_{k+1}=T,~~ t\in (t^{n}_{r},t^{n}_{r+1}],
\ed
such that $\max_{i}(t^{n}_{i+1}-t^{n}_{i})\to 0$ as $n\to \infty$)
$Y_{n}(t)$ converging to $Y(t)$ in probability.
The It\^o integral $\int_{0}^{t} Y_{n}(u) \ud W(u)$ is defined as,
{\small
\bd
\begin{array}{l}
\int\limits_{0}^{t} Y_{n}(u)  \ud W(u) :=
\sum\limits_{i=1}^{r} Y_{n}(t_{i-1})\left(W(t_{i})-W(t_{i-1})\right)+
Y_{n}(t_{r})\left(W(t)-W(t_{r-1})\right).
\end{array}
\ed
}
It can be shown that the sequence $Y_{n}$ of processes exists \citep{FKle,BOxe}. 

Adopting the  convention that matrices are written in bold--face, vectors as columns in normal font
with an arrow above \mbox{
($\vec{Z}$)} and scalar values in normal font, 
systems of stochastic differential equations can be written similarly,
\be\label{eqmvsde}
\ud \vec{X}(t) = \vec{\mu}(t,\vec{X}(t))\ud t + \mathbf{\Sigma}(t,\vec{X}(t)) \ud \vec{W}(t)),
\ee
with initial condition $\vec{X}_{0}$. The difference is that now 
$\vec{X}(t)$, $\vec{\mu}(t,\vec{X}(t))$ are vector valued process, 
$\mathbf{\Sigma}(t,\vec{X}(t))$ is a matrix valued process and
$\vec{W}(t)$ is a multidimensional standard Wiener processes.
As before Eq. \eqref{eqmvsde} is interpreted as,
\be\label{eqmvsdeIto}
\vec{X}(t) = \vec{X}(0) + \int\limits_{0}^{t} \vec{\mu}(u,\vec{X}(u))\ud u + \int\limits_{0}^{t} \mathbf{\Sigma}(u,\vec{X}(u)) \ud \vec{W}(u)),
\ee
where we understand $\int_{0}^{t} \vec{Y}(u) \ud \vec{W}(u)$ as a vector of entry--wise It\^o stochastic integrals. 

A natural question to ask is whether Eqs. (\ref{eqsde}, \ref{eqmvsde}, \ref{eqsdeIto}, \ref{eqmvsdeIto})
uniquely define a process (solution to the SDE) and does this process (solution) exist. The sufficient condition
\citep{FKle} for this is 
that the functions $\vec{\mu}(\cdot,\cdot)$ and $\mathbf{\Sigma}(\cdot,\cdot)$ satisfy a local Lipshitz condition,
\bd
\begin{array}{l}
\forall_{m \in \mathbb{N}} \exists_{K_{m} > 0} \forall_{t \in [0,T]} \forall_{\vec{x}, \vec{y} \in \mathbb{R}^{d_{X}}} 
\Bigg\{\Vert \vec{x} \Vert, \Vert \vec{y} \Vert  \le m 
\\ \Longrightarrow 
\Vert \vec{\mu}(t,\vec{x}) - \vec{\mu}(t,\vec{y}) \Vert^{2} + \sum\limits_{i=1}^{d_{X}}\sum\limits_{j=1}^{d_{W}} \left(\mathbf{\Sigma}_{i,j}(t, \vec{x})-\mathbf{\Sigma}_{i,j}(t, \vec{y})\right)^{2} \le 
K_{m} \Vert \vec{x} - \vec{y}\Vert^{2} \Bigg\}
\end{array}
\ed
and a global linear growth condition,
\bd
\exists_{C>0} \forall_{t\in [0,T]} \forall_{\vec{x} \in \mathbb{R}^{d_{X}}}
\Vert \vec{\mu}(t,\vec{x}) \Vert^{2} + \sum\limits_{i=1}^{d_{X}}\sum_{j=1}^{d_{W}} \mathbf{\Sigma}_{i,j}(t, \vec{x})^{2} \le C(1+\Vert \vec{x}\Vert^{2}),
\ed
where
$\Vert \cdot \Vert$ is the Euclidean norm, $d_{X}$ the dimension of $\vec{X}$ and
$d_{W}$ the dimension of $\vec{W}$. In the univariate case the 
conditions are the same with $d_{X}=d_{W}=1$.

The above has of course been a very compact summary of the bare essentials of how we understand
stochastic differential equations. In our modelling frameworks we do not focus on the mathematical theory
of stochastic differential equations but rather exploit the Markovian nature and finite dimensional
distributions of the solution (for very specific ones --- the Brownian motion and Ornstein--Uhlenbeck/Va\v si\v cek process).
A mathematical introduction to stochastic differential equations can be read from
e.g. \citet[][]{FKle,PMed,BOxe,SIac}. Additionally \citet{SIac} provides
rich R code concerning stochastic differential equations.

\subsection{Conditioned branching processes}
The second part of a tree--free model is the
model of the phylogenetic tree. We choose the framework of \emph{conditioned branching processes}
(conditioned --- on number of contemporary species). This topic has received significant attention
in the recent years
\citep[e.g.][]{DAldLPop2005,TGer2008a,TSta2009,TSta2011b,ALam2010,TStaMSte2012,AMooOGasTStaHLiMSte2012}.
The basic mathematical object is a branching process with 
constant birth and death rates (respectively $\lambda \ge \mu \ge 0$). 
The intuitive construction of such a process is that we start with
a single particle at time $0$ and then the particle either lives for an exponential (with rate $\mu$)
time and dies or lives for an exponential (with rate $\lambda$) time and  splits into two
particles behaving identically as the parent. From our perspective, where the branching process
is a modelling tool, this informal definition is sufficient. A formal treatment can be found in 
e.g. \citet{THar1963,KAthPNey2000} while \citet{MKimDAxe2002,PHacPJagVVat2007} discuss
combining branching processes with biological questions.

We are interested in conditioning on $n$, the number of current species represented by evolving particles
of our branching processes. The motivation behind this in the phylogenetic comparative
methods setting is obvious. We have our comparative sample 
and we assume that we have measured all species in the clade. We note that this is of course
a simplification and one should rather consider a framework (and this is one of the possible
future directions of developing this thesis) where the number of observed
species is just a random sample from the true (possibly unknown) number of species. 
It has already been mentioned in the introduction this can be an issue especially when studying
organisms lower than simple invertebrates. 

\citet{TGer2008a} presents the important properties of conditioned branching
processes. A key property is that conditional on the tree height the times to coalescent
of extant nodes are independent, identically distributed random variables. 
As we don't know the tree height, only the number of contemporary nodes, 
the strategy is to introduce a prior (before observing the number of extant nodes) distribution for it.
Using this prior and the above mentioned independence property we obtain posterior
distributions of characteristics of interest e.g. tree height, times of speciation events.
\citet{DAldLPop2005,TGer2008a} propose to use an (improper) uniform prior distribution on $(0,\infty)$ 
for the tree height. In the case of the pure birth tree ($\lambda>\mu=0$) this is equivalent
\citep{KHarDWonTSta2010,TStaMSte2012}
to stopping the tree just before the $n$--th speciation event (i.e. just before there are
$n+1$ species). Additionally the interspeciation times in a Yule tree can be elegantly
characterized. Namely the time between the $(k-1)$--th  and $k$--th speciation events 
are exponentially distributed with rate $k\lambda$ as this is the minimum of $k$ rate
$\lambda$ exponential random variables \citep[][and Papers IV, V, VI]{WFel1971,TGer2008b}.

\subsection{Properties of the gamma function}
The gamma function appears in formulae for important properties of our tree--free models.
Below we summarize the well known (and exploited by us) properties of this function.
The gamma function is defined as,
\bd
\Gamma(z) = \int\limits_{0}^{\infty} t^{z-1}e^{-t}\ud t
\ed
and the function has the fundamental recursive property 
$\Gamma(z+1)=z\Gamma(z)$ resulting for integer $z$, $\Gamma(z)=(z-1)!$. There is a close
relation between the gamma function and the beta function $B(x,y)$,
\bd
B(x,y) = \int\limits_{0}^{1} t^{x-1}(1-t)^{y-1}\ud t = \frac{\Gamma(x)\Gamma(y)}{\Gamma(x+y)}
\ed
and for large $x$ and fixed $y$, $B(x,y) \sim \Gamma(y)x^{-y}$.
In Papers IV and V we rely heavily on the below summation formula
(verifiable by induction if $z\neq y$ and directly if $z=y$),
\be
\sum\limits_{k=1}^{n-1}\frac{\Gamma(k+y)}{\Gamma(k+z+1)} = 
\left\{
\begin{array}{ll}
\frac{\Gamma(n+z)\Gamma(y+1)-\Gamma(z+1)\Gamma(n+y)}{\Gamma(z+1)\Gamma(n+z)(z-y}, & z \neq y, \\
\Psi(n+y)-\Psi(1+y), & z=y,
\end{array}
\right.
\ee
where $\Psi(z)$ is the polygamma function defined as $\Psi(z)=\Gamma'(z)/\Gamma(z)$.
This function has the property that for natural $z>0$, $\Psi(z)=\gamma + H_{z-1,1}$, where
$\gamma\approx 0.577$ is the Euler--Mascheroni constant
and
$H_{n,k}$ is the $k$--th generalized harmonic number,
\bd
H_{n,k} = \sum\limits_{i=1}^{n}\frac{1}{i^{k}}
\ed
(instead of $H_{n,1}$ in Paper III $a_{n}$ is used, in Paper IV $h_{n}$, in Paper V $H_{n}$).

\clearpage
\newpage

\section{Interacting traits}
\subsection{Paper I --- Measurement error in Phylogenetic Comparative Methods}\label{sbsecPapI}
Paper I concerns the issue of correcting for measurement error 
in regression where the predictor variables and measurement errors can
be correlated in an arbitrary way. 
The bias caused by measurement error in regression is a well studied
topic in the case of independent observations \citep[see e.g.][]{WFul1987,JBuo}. 
However as we show in 
\mbox{Paper I} everything complicates when the predictor variables
are dependent. If in addition we allow for measurement errors to be dependent
between observations we get an even more complicated situation. 
This is typically the situation with comparative data. The predictor 
variables are usually species' traits evolving on the phylogeny.
The phenotypic data is commonly species'
means from a number of observations. Attached to these means is the
intra--species variability, being statistically the same as (phylogenetically) dependent measurement errors.

We adopt the following
notation: a variable with subscript $_{o}$ will denote an observed (with error)
variable, while the subscript $_{t}$ will mean the true, unobserved value of the variable.
Additionally 
by $\mathbf{I}$ we denote the identity matrix of appropriate size. For a matrix $\mathbf{M}$, $\mathbf{M}^{T}$
denotes
the transpose of $\mathbf{M}$ and $\mathbf{M}^{-1}$ the inverse.
The operation $\mathrm{vec}(\mathbf{M})$ means
the vectorization, \emph{i.e.} stacking of columns onto each other, of the
matrix $\mathbf{M}$ and $\mathrm{vec}^{-1}(\vec{Z})$ the inverse of $\mathrm{vec}$ operation
(assuming matrix sizes are known). 
For two random variables $X$ and $Y$ the notation $X \independent Y$ means that they
are independent.

We consider the general linear model,
\be\label{eqGLM}
\begin{array}{rclc}
\vec{Y}_{t} & = & \mathbf{D}_{t}\vec{\beta} + \vec{r}_{t}, & \vec{r}_{t} \sim \mathcal{N}(\vec{0},\mathbf{V}_{t}),
\end{array}
\ee
where $\vec{Y}_{t}$ is a vector of $n$ observations of the dependent variable, $\vec{\beta}$ is a
vector of $m$ parameters to be estimated, $\mathbf{D}_{t}$ is an $n \times m$ design matrix 
and $\vec{r}_{t}$ is a vector of $n$ noise terms with $n\times n$ covariance
matrix $\mathbf{V}_{t}$. 

To the general linear model of Eq. \eqref{eqGLM} we want to introduce a measurement error model. 
We write the model with errors in the design matrix and the response variables as,
\bd
\begin{array}{rcl}
\mathbf{D}_{o} & = & \mathbf{D}_{t}+ \mathbf{U}, \\
\vec{Y}_{o} & = & \vec{Y}_{t}+ \vec{e}_{y},
\end{array}
\ed
where $\mathbf{U}$ is a $n\times m$ matrix of random observation errors in the elements of $\mathbf{D}_{t}$,
and $\vec{e}_{y}$ is a vector of length $n$ of observation errors in $\vec{Y}_{t}$. Each column
of $\mathbf{U}$ is a vector of observation errors for a predictor variable. Furthermore
we assume that $\mathrm{vec}(\mathbf{D}_{t})$ and $\mathrm{vec}(\mathbf{U})$ are 
zero--mean normal random vectors, with covariance matrices
equalling $\mathbf{V}_{d}$ and $\mathbf{V}_{u}$ respectively. 
Putting all the elements together,  
the regression with observation error model is the following,
\bd
\begin{array}{rclcc}
\vec{Y}_{t} & = & \mathbf{D}_{t}\vec{\beta} + \vec{r}_{t}, 
& \mathrm{vec}(\mathbf{D}_{t}) \sim \mathcal{N}(\vec{0},\mathbf{V}_{d}), & \vec{r}_{t} \sim \mathcal{N}(\vec{0},\mathbf{V}_{t}), \\
\vec{Y}_{o} & = & \vec{Y}_{t} + \vec{e}_{y}, & \vec{e}_{y} \sim \mathcal{N}(\vec{0},\mathbf{V}_{e}), \\
\mathbf{D}_{o} & = & \mathbf{D}_{t} + \mathbf{U}, & \mathrm{vec}(\mathbf{U}) \sim \mathcal{N}(\vec{0},\mathbf{V}_{u}), \\
\vec{Y}_{o} & = & (\mathbf{D}_{o}-\mathbf{U})\vec{\beta} + \vec{r}_{t}+\vec{e}_{y}. 
\end{array}
\ed

If we further assume that all the errors are independent of each other and the other variables,
i.e. $\vec{r}_{t} \independent \vec{e}_{y}$, $\vec{r}_{t} \independent \mathbf{U}$,
$\mathbf{U} \independent \vec{e}_{y}$, $\mathbf{U} \independent \mathbf{D}_{t}$ and 
$\vec{Y}_{t} \independent \vec{e}_{y}$ then we can write the last equation as, 
$\vec{Y}_{o}  =  (\mathbf{D}_{o}-\mathbf{U})\vec{\beta} + \vec{r}$,
where $\vec{r}=\vec{r}_{t}+\vec{e}_{y}$, so
$\vec{r} \sim \mathcal{N}(\vec{0},\mathbf{V})$,
where $\mathbf{V}=\mathbf{V}_{t}+\mathbf{V}_{e}$.
In Paper I we represent the noise as $\vec{r}=\vec{r}_{t}+\vec{e}_{y} - \mathbf{U}\vec{\beta}$
and consider the regression model $\vec{Y}_{o}  =  \mathbf{D}_{o}\vec{\beta} + \vec{r}$ but 
here we do not do this so that we do not have to consider iterative procedures.
We do not assume any structure (in particular diagonality) of the covariance matrices $\mathbf{V}$, $\mathbf{V}_{t}$, 
$\mathbf{V}_{e}$, $\mathbf{V}_{d}$ or $\mathbf{V}_{u}$. 
The generalized least squares estimator of $\vec{\beta}$ can be written, as
\bd
\begin{array}{rcl}  
\hat{\vec{\beta}}&=&(\mathbf{D}_{o}^{T}\mathbf{V}^{-1}\mathbf{D}_{o})^{-1}\mathbf{D}_{o}^{T}\mathbf{V}^{-1}((\mathbf{D}_{o}-\mathbf{U})\vec{\beta}+\vec{r})
\\&=&
(\mathbf{D}_{o}^{T}\mathbf{V}^{-1}\mathbf{D}_{o})^{-1}\mathbf{D}_{o}^{T}\mathbf{V}^{-1}(\mathbf{D}_{t}\vec{\beta}+\vec{r}).
\end{array}
\ed

We want to compute the expectation of this estimator conditional on the observed predictor variables $\mathbf{D}_{o}$.
We assumed $\vec{r}$ had zero mean and is independent of $\mathbf{D}_{o}$ so we have,
\bd
\begin{array}{rcl}
\E{\hat{\vec{\beta}} \vert \mathbf{D}_{o}} 
&=&
(\mathbf{D}_{o}^{T}\mathbf{V}^{-1}\mathbf{D}_{o})^{-1}\mathbf{D}_{o}^{T}\mathbf{V}^{-1}\E{\mathbf{D}_{t} \vert \mathbf{D}_{o}}\vec{\beta} 
\\ &=&
(\mathbf{D}_{o}^{T}\mathbf{V}^{-1}\mathbf{D}_{o})^{-1}\mathbf{D}_{o}^{T}\mathbf{V}^{-1}\E{\mathbf{D}_{o} -\mathbf{U} \vert \mathbf{D}_{o}}\vec{\beta}
\\ &=&
(\mathbf{I}-(\mathbf{D}_{o}^{T}\mathbf{V}^{-1}\mathbf{D}_{o})^{-1}\mathbf{D}_{o}^{T}\mathbf{V}^{-1}\E{\mathbf{U} \vert \mathbf{D}_{o}})\vec{\beta}.
\end{array}
\ed

It remains to calculate $\E{\mathbf{U} \vert \mathbf{D}_{o}}$, 
to do this we need to work with its vectorized form, 
$\E{\mathrm{vec}(\mathbf{U}) \vert \mathbf{D}_{o}}$.
Using common facts about the multivariate normal distribution
and that $\mathbf{D}_{o}=\mathbf{D}_{t}+\mathbf{U}$,
\bd
\begin{array}{l}
\E{\mathrm{vec}(\mathbf{U}) \vert \mathbf{D}_{o}} \\
=\E{\mathrm{vec}(\mathbf{U})}+\cov{\mathrm{vec}(\mathbf{U})}{\mathrm{vec}(\mathbf{D}_{o})}
\Cov{\mathrm{vec}(\mathbf{D}_{o})}^{-1}(\mathrm{vec}(\mathbf{D}_{o})-\E{\mathrm{vec}(\mathbf{D}_{o})})\\
=\Cov{\mathrm{vec}(\mathbf{U})}\Cov{\mathrm{vec}(\mathbf{D}_{o})}^{-1}\mathrm{vec}(\mathbf{D}_{o})
=\mathbf{V}_{u}\mathbf{V}_{o}^{-1}\mathrm{vec}(\mathbf{D}_{o}),
\end{array}
\ed
where $\mathbf{V}_{o}=\mathbf{V}_{d}+\mathbf{V}_{u}$.
Therefore we have that,
\be
\E{\hat{\vec{\beta}} \vert \mathbf{D}_{o}} =
\mathbf{K}\vec{\beta},
\label{eqBias}
\ee
where the matrix $\mathbf{K}$ (reliability matrix) is given by
\be
\mathbf{K} =
\mathbf{I}-(\mathbf{D}_{o}^{T}\mathbf{V}^{-1}\mathbf{D}_{o})^{-1}\mathbf{D}_{o}^{T}\mathbf{V}^{-1}
\mathrm{vec}^{-1}(\mathbf{V}_{u}\mathbf{V}_{o}^{-1}\mathrm{vec}(\mathbf{D}_{o}).
\label{eqK}
\ee

The above is a general formula for 
$\mathbf{V}_{o}$ and $\mathbf{V}_{u}$ of any form.
Imposing some structure on them can simplify the formula for $\mathbf{K}$
and we can notice that no assumptions are needed on the covariance matrices $\mathbf{V}_{t}$ 
and $\mathbf{V}_{e}$. 
Below we consider some special 
cases of $\mathbf{V}_{o}$ and $\mathbf{V}_{u}$.
Other generalizations and including fixed effects are discussed by
\citet{KBarLic2011}.
\\
\emph{Independent observations of predictors} where $\mathbf{\Sigma}_{d}$ is the covariance 
matrix of the true (unobserved) predictors in a given observation and $\mathbf{\Sigma}_{u}$
is the covariance matrix of the measurement errors in predictors in a given observation,
\be\label{eqMultIndepBias}
\mathbf{K} = (\mathbf{\Sigma}_{d}+\mathbf{\Sigma}_{u})^{-1}\mathbf{\Sigma}_{d}.
\ee
\\
\emph{Independent predictors} (i.e. the predictors and their errors are independent
between each other but not between observations), now
$\mathbf{\Sigma}_{d_{i}}$ is the covariance 
matrix of the true (unobserved) $i$th predictor between observations and $\mathbf{\Sigma}_{u_{i}}$
the covariance matrix of the measurement error and $\mathbf{d}_{o_{i}}$ denotes the $i$th column of $\mathbf{D}_{o}$,
\be\label{eqMultIndepPredBias}
\mathbf{K} = \mathbf{I}-
(\mathbf{D}_{o}^{T}\mathbf{V}^{-1}\mathbf{D}_{o})^{-1}\mathbf{D}_{o}^{T}\mathbf{V}^{-1}
\left[\mathbf{\Sigma}_{u_{1}}\mathbf{\Sigma}_{o_{1}}^{-1}\mathbf{d}_{o_{1}};
\ldots ;\mathbf{\Sigma}_{u_{m}}\mathbf{\Sigma}_{o_{m}}^{-1}\mathbf{d}_{o_{m}}
\right].
\ee
\\
\emph{Single predictor}
\be\label{eqSingBias}
\mathbf{K} = \mathbf{I}-(\mathbf{D}_{o}^{T}\mathbf{V}^{-1}\mathbf{D}_{o})^{-1}\mathbf{D}_{o}^{T}\mathbf{V}^{-1}\mathbf{V}_{u}\mathbf{V}_{o}^{-1}\mathbf{D}_{o}
\ee

In the case of a single predictor with independent observations 
it is well known that $\kappa \in (-1,1)$ (one--dimensional counterpart of Eq. \ref{eqMultIndepBias}).
Therefore the bias corrected estimator has a larger variance than the uncorrected one. This immediately
implies a trade--off if we use the mean square error, $\E{(\hat{\beta}-\beta)^{2}}$, 
to compare estimators. 

In Paper I we study the case of dependent 
observations of a single predictor. We transform the formulae for the mean square
errors of both estimators into a criterion in terms of estimable quantities,
$$\sigma^{2}_{\beta}/\kappa^{2} < \sigma^{2}_{\beta}+(\kappa-1)^{2}\beta^{2},$$
where $\sigma^{2}_{\beta}$ is the variance of the estimator of $\beta$.
Unlike in the independent observations case $\kappa$ can take values outside
the interval $(-1,1)$. 
\subsection{Paper II --- Multivariate Phylogenetic Comparative Methods}\label{sbsecPapII}
Due to the increase in computational power there has been a recent development of 
phylogenetic Ornstein--Uhlenbeck models 
\citep[e.g.][]{MButAKinOUCH,THanJPieSOrzSLOUCH,ALabJPieTHan2009,JBeaDJhwCBoeBOMe2012,TIngDMah2013}. 
Paper II is a novel addition to this direction and considers  
an adaptive multivariate 
stochastic differential equation model of phenotype evolution.
It includes an R software package,
mvSLOUCH (\textit{m}ulti\textit{v}ariate \textit{S}tochastic \textit{L}inear \textit{O}rnstein--\textit{U}hlen\-beck 
models for phylogenetic \textit{C}omparative \textit{H}ypotheses)
covering nearly all cases of \citet{EMarTHan96,THan1997,EMarTHan97}'s framework
(except some cases where the drift matrix is singular or does not have an eigendecomposition).

The first modelling approach via stochastic differential equations for 
comparative methods in a continuous trait setting is due to \citet{JFel1985}. 
He proposed that the phenotype evolves as a Brownian motion along the phylogeny. 
A trait $X$ is evolving as a Brownian motion if it can be described by the following 
stochastic differential equation along a single lineage,
\be
\ud X(t) =  \sigma \ud W(t),
\ee
where $\ud W(t)$ is white noise. 
This means that the displacement of the trait after time $t$ from its initial 
value is $X(t) \sim \mathcal{N}(X(0),\sigma^{2} t)$. This model assumes that there is absolutely no selective pressure
on trait $X$ whatsoever, so that the trait value just randomly fluctuates from its initial starting point. 

A multi--dimensional Brownian motion model is immediate,
\be
\ud \vec{X}(t) = \mathbf{\Sigma} \ud \vec{W}(t),
\ee
where $\mathbf{\Sigma}$ will be the diffusion matrix and $\vec{W}$ will be 
the multidimensional Wiener process. All components of the random vector are normally distributed as 
$\vec{W}(t)-\vec{W}(s) \sim \mathcal{N}(\vec{0},(t-s)\mathbf{I})$ and so
$\vec{X}(t) \sim \mathcal{N}(\vec{X}(0), t\mathbf{\Sigma}\mathbf{\Sigma}^{T})$. 
The covariance matrix of all of the phylogenetically dependent data will be 
$\mathbf{V}=\mathbf{T}\otimes (\mathbf{\Sigma}\mathbf{\Sigma}^{T})$, where
$\otimes$ is the Kronecker product and 
$\mathbf{T}$ is the matrix of divergence times (time from the origin of the tree to the point of divergence) 
of species on the phylogeny.
Such Brownian trait evolution has the property of time reversibility
so both $\vec{X}(t) - \vec{X}(0)$ and $\vec{X}(0) - \vec{X}(t)$ 
will be identically distributed. Using
this one can find an independent sample on the phylogenetic tree, i.e. 
the contrasts between nodes such that
no pair of nodes shares a branch in the path connecting them. 
This observation was exploited by \citet{JFel1985} in his  independent 
contrasts estimation algorithm \citep[but see also][]{RFre2012}.

The lack of the possibility of adaptation in the Brownian Motion model was
discussed by \citet{THan1997,THanSOrz2005} and therefore a more complex 
Ornstein--Uhlenbeck type of framework was proposed,
\be
\begin{array}{rcl}\label{eqOU}
\ud \vec{Z} (t) & = & -\mathbf{F}(\vec{Z}(t)-\vec{\Psi}(t))\ud t + \mathbf{\Sigma} \ud \vec{W}(t),
\end{array}
\ee
termed here the Ornstein--Uhlenbeck model. The $\mathbf{F}$ matrix is called the drift matrix, $\vec{\Psi}(t)$
the drift vector and $\mathbf{\Sigma}$ the diffusion matrix. 

In Paper II we  consider in detail a very important decomposition 
of $\mathbf{F}$ ---
the multivariate Ornstein--Uhlenbeck Brownian motion (mvOUBM) model
\citep[a multivariate generalization of the OUBM model due to][]{THanJPieSOrzSLOUCH}
\be\label{eqGenModel2}
\begin{array}{c}
\ud \left[\begin{array}{c} \vec{Y} \\ \vec{X} \end{array}\right] (t) 
=  -\left[ \begin{array}{c|c} \mathbf{A} & \mathbf{B} \\ \hline \mathbf{0} & \mathbf{0} \end{array} \right]
\left(\left[\begin{array}{c} \vec{Y} \\ \vec{X} \end{array} \right] (t) -
\left[\begin{array}{c} \vec{\psi} \\ \vec{0} \end{array} \right] (t) \right)\ud t 
\\ +
\left[\begin{array}{c|c} \mathbf{\Sigma}^{yy} & \mathbf{\Sigma}^{yx} \\ \hline \mathbf{\Sigma}^{xy} & \mathbf{\Sigma}^{xx} \end{array} \right]
\ud \mathbf{W}(t).
\end{array}
\ee
 
Assuming that $\mathbf{F}$ and $\mathbf{A}$ have positive real part eigenvalues then respectively
the $\vec{\Psi}(t)$ function represents a deterministic optimum value for $\vec{Z}(t)$, Eq. \eqref{eqOU}
and $\vec{\psi}(t)$ the deterministic part of the optimum for $\vec{Y}(t)$ in Eq. \eqref{eqGenModel2}.
If we write out the SDEs in vector form then
$f_{ij}/f_{ii}$ and $a_{ij}/a_{ii}$, where $f_{ij}$ are the elements of $\mathbf{F}$ and $a_{ij}$ of $\mathbf{A}$,
can be understood as effects of $Z_{j}$ and $Y_{j}$ on the primary optimum
of $Z_{i}$ and $Y_{i}$ respectively \citep{THan1997,THanJPieSOrzSLOUCH}. 
The eigenvalues, if they have positive 
real part, can be understood to control the speed of the traits' approach to their
optima, while the eigenvectors indicate the path towards the optimum.
The diffusion matrix in general
represents stochastic perturbations to the traits' approach to their optima. These
perturbations can come from many sources internal or external, e.g.
they can represent unknown/unmeasured components of the system under study
or some random genetic changes linked to the traits.
 
Paper II includes an R package, mvSLOUCH, that implements a 
(heuristic) maximum likelihood estimation method to estimate the parameters
of the stochastic differential equation \eqref{eqOU}. The package
also allows for the estimation of parameters of the special
submodel defined by Eq. \eqref{eqGenModel2}. 
The implementation of the estimation procedure requires
combining advanced linear algebra and computing techniques 
with probabilistic and statistical results. These have been discussed in detail by
\citet{KBarLic2011}.

Paper II also includes a reanalysis of a Cervidae data set
\citep[deer antler length, male and female body masses, compiled by][]{FPlaCBonJGai}.
Our results generally
confirm those of \cite{FPlaCBonJGai}, that (i) 
there is positive linear relationship between the logarithms of antler length and male body mass
and (ii) the mating tactic does not influence directly the antler length
nor the male body mass. The allometry between antler length and
male body mass is greater than $1/3$ indicating that there is more than just 
a proportional increase of antler length (1--dimensional)
when body mass increases (3--dimensional). We can also observe
that the estimates of the effect of breeding group size
on the antler length and male body mass are larger with the
increase in breeding group size. 

However our analysis in addition shows
that antler length and male body mass are adapting very rapidly
to changes in female body mass. We can also see that they
adapt independently of each other, there is no direct influence of
one variable on the primary optimum of the other and all dependencies
between antler length and male body mass are due to the common female
body mass predictor variable, and correlations in noise pushing them 
away from their respective optima. 
This adaptation result was not observed by
\citet{FPlaCBonJGai} as they only considered a Brownian motion 
evolutionary model.
\clearpage
\newpage
\section{Tree--free models}
\subsection{Paper III --- Interspecies correlation for neutrally evolving traits}\label{sbsecPapIII}
Paper III is a first step to combine branching process with stochastic models of
phenotype evolution. We  asked how similar 
do we expect species phenotypes to be.

To measure the phenotypic similarity we concentrate on the interspecies correlation coefficient,
defined classically as,
\be \label{eqrhon1}
\rho_{n} = \frac{\cov{X_{1}}{X_{2}}}{\sqrt{\var{X_{1}}\var{X_{2}}}},
\ee
where $X_{1}$ and $X_{2}$ are a randomly sampled pair of tip species. We also assume
that all $n$ species are contemporary,
so Eq. \eqref{eqrhon1} will simplify,
\be \label{eqrhon2}
\rho_{n} = \frac{\cov{X_{1}}{X_{2}}}{\var{X}},
\ee
where $X$ is a randomly sampled tip species. 
We have two sources of variation the randomness in the tree and the phenotypic evolution. 
As the considered phenotypic processes are Markovian and the phenotype is assumed not to
influence the branching then using the laws of total variance and covariance we have,
\be\label{eqvarcovYBM}
\begin{array}{rcl}
\var{X} & = & \E{\var{X \vert U_{n}}} + \var{\E{X \vert U_{n}}}, \\
\cov{X_{1}}{X_{2}} & = & \E{\cov{X_{1}}{X_{2} \vert U_{n}, \tau^{(n)}}} 
\\ && + \cov{\E{X_{1} \vert U_{n}, \tau^{(n)}}}{\E{X_{2} \vert U_{n}, \tau^{(n)}}}
\\ &=&
\E{\cov{X_{1}}{X_{2} \vert U_{n}}} + \var{\E{X_{a_{12}} \vert U_{n},\tau^{(n)}}},
\end{array}
\ee
where $X_{a_{12}}$ is the most recent common ancestor of the two randomly
sampled species, 
$U_{n}$ ($T$ in Papers III, V) is the height of the tree and
$\tau^{(n)}$ ($\tau$ in Papers III, V) is the time to coalescent of a random
pair of tip species.  To calculate these values we need to 
assume a stochastic model for the phenotype evolution and for the tree.
In Paper III we assumed the Brownian motion model. 
The Brownian motion model is characterized by two parameters
$X_{0}$ the ancestral state and $\sigma$ the diffusion coefficient.

Under it the conditional (on tree) mean, variance and covariance are,
\be
\begin{array}{rcl}
\E{X \vert U_{n}} & = & X_{0}, \\
\var{X \vert U_{n}} & = & \sigma^{2}U_{n}, \\ 
\cov{X_{1}}{X_{2} \vert U_{n}, \tau^{(n)}} & = & \sigma^{2}(U_{n}-\tau^{(n)}).
\end{array}
\ee
From Eq. \eqref{eqvarcovYBM} we see that we need to know
the expectation of $U_{n}$ and $(U_{n}-\tau^{(n)})$ --- these are dependent on the assumed model of phylogeny.
Our phylogeny modelling approach is heavily based on the findings of T. Stadler (ne\'e Gernhard)
concerning conditioned branching processes.
The phylogenetic tree is modelled as a constant rate birth--death process
($\lambda$, $\mu$ being the birth and death rates) 
conditioned on $n$ contemporary tips. 
We consider three regimes,
$\mu=0$ (Yule, pure--birth case), $0<\mu<\lambda$ (supercritical case),
$\mu=\lambda$ (critical case) and in each case calculate the expectations,
obtaining the following results in the pure--birth \mbox{
($0=\mu< \lambda$)} case,
\be
\begin{array}{rcl}
\E{U_{n}} & = & \frac{1}{\lambda}H_{n,1}, \\
\E{U_{n}-\tau^{(n)}} & = & \frac{2}{\lambda(n-1)}\left(n-H_{n,1} \right),
\end{array}
\ee
and supercritical ($0<\mu<\lambda$) case,
\be
\begin{array}{rcl}
\E{U_{n}} & = & \frac{1}{\mu(\lambda/\mu-1)}\left(H_{n,1}+e_{n,\lambda/\mu}-\ln\frac{\lambda/\mu}{\lambda/\mu-1} \right),  \\
\E{U_{n}-\tau^{(n)}} & = &\frac{2\left(n+ne_{n,\lambda/\mu}-\frac{\lambda/\mu}{\lambda/\mu-1}\left(H_{n,1}+e_{n,\lambda/\mu}-\ln\frac{\lambda/\mu}{\lambda/\mu-1} \right) \right)}{\mu(\lambda/\mu-1)(n-1)},
\end{array}
\ee
where, $e_{n,y}=\int_{0}^{1}x^{n}/(y-x)\ud x$.
Taking the quotient gives the correlation coefficient, decaying to $0$ in both cases as $2/\ln n$.
The third critical case needs more careful treatment as here $\E{U_{n}} = \infty$.
Therefore the variance is undefined. However we define a proper prior distribution
for $U_{n}$, in Paper III $U_{n}\sim \mathrm{Uniform}[0,N]$, 
calculate the necessary expectations ($m:=(N+1)/N$),
{\small
\be
\begin{array}{rcl}
\E{U_{n}} & = & ne_{n,m}\\
\E{U_{n}-\tau^{(n)}} & = & 2ne_{n,m}-\frac{2}{n-1}\left(\frac{ne_{n,m}+n}{m-1}-\frac{m}{(m-1)^{2}}\left(H_{n,1}+e_{n,m}-\ln \frac{m}{m-1} \right) \right),
\end{array}
\ee
}and study the interspecies correlation coefficient 
for two asymptotic regimes,
\begin{itemize}
\item {\small
$\rho_{n} = 1-\frac{1}{2(\ln N - H_{n,1})+o(1)}$ as $N \to \infty$}
\item {\small
$\rho_{n} \to 2 - \frac{2}{\alpha}\left(1+\frac{1}{e^{\alpha}Ei(\alpha)} \right) + \frac{2}{\alpha^{2}}\left(1+\frac{\ln \alpha + \gamma}{e^{\alpha}Ei(\alpha)} \right)$ as 
}
\mbox{$N,n \to \infty$, $n/N \to \alpha$}, $Ei(z)$ is the exponential integral
and $\gamma$ is the Euler--Mascheroni constant.
\end{itemize}

In addition to the correlation coefficient we are able to calculate two formulae useful from 
an estimation point of view --- the variance of the sample mean and expectation of the sample
variance. Both can be written in terms of the interspecies correlation coefficient,
\be
\begin{array}{rcl}
\var{\overline{X}_{n}} & = & \frac{1+(n-1)\rho_{n}}{n}\var{X},\\
\E{S_{n}^{2}} & = & \left(1-\rho_{n} \right)\var{X}
\end{array}
\ee
and in the Brownian motion case they equal for the different regimes,
$(0=\mu< \lambda)$, $(0<\mu< \lambda)$, $(0<\mu=\lambda)$ respectively,
{\small
\be
\begin{array}{rcl}
\var{\overline{X}_{n}} & = & \sigma^{2}\left\{
\begin{array}{l}
\frac{1}{\lambda}\left(2-\frac{H_{n,1}}{n}\right), \\
\frac{2}{\mu}\left(\frac{1+e_{n,\lambda/\mu}}{\lambda/\mu-1}-\frac{1}{n}\frac{\lambda/\mu+1}{n(\lambda/\mu-1)^{2}}\left(H_{n,1}+e_{n,\lambda/\mu}-\ln\frac{\lambda/\mu}{\lambda/\mu-1} \right) \right), \\
\frac{1}{\lambda}\left(e_{n,m}\left(2n-2N+\frac{2N(N+1)}{n}-1 \right)-\frac{A_{n,N}}{n-1} \right),
\end{array}
\right.
\end{array}
\ee
}
where $A_{n,N}=2N(n-(N+1)(H_{n,1}-\ln(N+1)))$ $\mathbf{\mu,\lambda}$ and
{\small
\be
\begin{array}{rcl}
\E{S_{n}^{2}} & = & \sigma^{2}\left\{
\begin{array}{l}
\frac{1}{\lambda}\left(\frac{n+1}{n-1}H_{n,1}-2\frac{n}{n-1} \right), \\
\frac{2}{\mu(n-1)(\lambda/\mu-1)}\left(\left(\frac{2\lambda/\mu}{\lambda/\mu-1}+n \right)\left(H_{n,1}-\ln\frac{\lambda/\mu}{\lambda/\mu-1}\right) \right.  \\
\left. +\left(\frac{2\lambda/\mu}{\lambda/\mu-1}-n\right)e_{m,\lambda/\mu}+1-2n \right), \\
\frac{1}{\lambda}\left(e_{n,m}\left(\frac{2Nn}{n-1}-\frac{2N(N+1)}{n-1}-n \right)+\frac{A_{n,N}}{n-1} \right),
\end{array}
\right.
\end{array}
\ee
}
with respective asymptotic behaviour, in the critical case $n/N\to\alpha$,
{\small
\be
\begin{array}{rcl}
\var{\overline{X}_{n}} & \sim & 
\sigma^{2}\left\{
\begin{array}{l}
\frac{2}{\lambda}, \\
\frac{2}{\mu(\lambda/\mu-1)}, \\
\frac{n}{\lambda}\left(\frac{2}{\alpha^{2}}\left(\alpha^{2}-\alpha+1 \right)e^{\alpha}Ei(\alpha)-\alpha+\ln\alpha\right),
\end{array}
\right.
\end{array}
\ee
\be
\begin{array}{rcl}
\E{S_{n}^{2}} & \sim & 
\sigma^{2}\left\{
\begin{array}{l}
\frac{\ln n}{\lambda} \\
\frac{\ln n}{\mu(\lambda/\mu-1)}, \\
\frac{n}{\lambda}\left(\frac{1}{\alpha^{2}}\left(2\alpha-\alpha^{2}+2 \right)e^{\alpha}Ei(\alpha)+\alpha-\ln\alpha\right).
\end{array}
\right.
\end{array}
\ee
} We can see that as $\var{\overline{X}_{n}} \to 2$ the sample average is not 
a consistent estimator of the ancestral state \citep[similarly obtained by][]{CAne}.
Knowledge of the variance of the sample mean can be developed into phylogenetic confidence intervals
for the ancestral state and based on the expectation of the sample variance we can 
construct an unbiased estimator for $\sigma^{2}$. In Paper IV we develop these concepts for the
Yule--Ornstein--Uhlenbeck process.

\subsection{Paper IV --- Tree--free phylogenetic confidence intervals for an adaptive model}\label{sbsecPapIV}
A natural continuation of Paper III was to consider the Ornstein--Uhlenbeck process
for phenotype evolution,
\be
\begin{array}{rcl}
\ud X(t) & = & -\alpha(X(t) - \theta) \ud t + \sigma \ud B(t).
\end{array}
\ee
In our framework we interpret that the trait $X(t)$ is adapting (we assume $\alpha>0$) and trying to approach
its adaptive peak (optimal value) --- $\theta$. We assume in Paper IV that the value of $\theta$ is constant
over the whole phylogeny but notice that this is not what is desired in 
comparative methods. In a common comparative methods setting we should be 
able to identify groups of species (and subclades of the phylogeny --- if available)
that have their own unique adaptive peaks.  
Therefore our results on the single adaptive--peak
could be useful for analyzing subclades and we leave a tree--free model with multiple adaptive
peaks as a possible development of Paper IV. 

In Paper IV we consider two classical estimators, the sample mean 
$$\overline{X}_{n}=\frac{1}{n}\sum\limits_{i=1}^{n}X_{i}$$
and the sample variance
$$S_{n}^{2} = \frac{1}{n-1}\sum\limits_{i=1}^{n}(X_{i}-\overline{X}_{n})^{2}. $$
It turns out that the sample mean is an asymptotically unbiased estimator
of $\theta$ and the sample variance is an asymptotically unbiased estimator 
of the stationary variance of the Ornstein--Uhlenbeck process, $\sigma^{2}/2\alpha$.
We also show that these estimators are consistent, $\var{\overline{X}_{n}},\var{S_{n}^{2}}\to 0$.
By this we are able to derive phylogenetic confidence intervals for $\theta$ 
that differ from the usual classical ones by a phylogenetic 
correction factor. This factor is greater than $1$ as we have a smaller effective sample size due to dependencies.
It turns out however that the form of these intervals (and of the variance of the sample
mean and sample variance) depends on the value of $\alpha$, with three possible regimes.
\begin{enumerate}[i)]
\item $\alpha > 0.5\lambda$ in this case (fast adaptation) the approximate confidence interval 
at level $1-x$ will be of the form 
\be
\overline{X}_{n} \pm z_{x/2}\frac{\sqrt{S_{n}^{2}}}{\sqrt{n}}\sqrt{\frac{2\alpha+1}{2\alpha-1}},
\ee
where $z_{x}$ is the $x$ level quantile of the standard normal distribution
and we also have a central limit theorem that the standardized sample mean
$(\overline{X}_{n}-\theta)/\sqrt{\sigma^{2}/2\alpha}$ is asymptotically mean--zero normally distributed
with variance $\frac{2\alpha+1}{2\alpha-1}$.
\item $\alpha=0.5\lambda$, there seems to be a phase transition for the adaptation rate equalling 
half the speciation rate. We can derive the following approximate phylogenetic confidence interval
\be
\overline{X}_{n} \pm z_{x/2}\frac{\sqrt{S_{n}^{2}}}{\sqrt{n}}\sqrt{2\ln n},
\ee
and the central limit theorem will hold for the following normalization of the sample average
$\sqrt{n/\ln n}(\overline{X}_{n}-\theta)/\sigma$ which will be asymptotically normal mean--zero with variance $2$.
\item $0<\alpha<0.5\lambda$, slow adaptation --- in this case the situation is more complicated,
the limiting distribution of the normalized \mbox{
$\frac{\overline{X}_{n}-\theta}{\sqrt{\sigma^{2}/2\alpha}}$} 
sample mean will depend on the starting position
and we cannot expect it to be normal \citep[compare to a very similar model due to][]{RAdaPMil20111,RAdaPMil20112}.
We can propose the following approximate confidence interval 
\be
\left(\overline{X}_{n}-q_{\alpha}(x/2)\sqrt{S^{2}_{n}}n^{\alpha}, \overline{X}_{n}+q_{\alpha}(1-x/2)\sqrt{S^{2}_{n}}n^{\alpha}\right)
\ee
but the quantiles, $q_{\alpha}(x/2)$, $q_{\alpha}(1-x/2)$, have to be currently obtained by simulation methods.
\end{enumerate}

The same sort of phase transition at $\alpha=0.5\lambda$ was noticed by \citet{RAdaPMil20111,RAdaPMil20112} in a 
related modelling setting. One way of thinking about this \citep[after][]{RAdaPMil20111,RAdaPMil20112}
is that for a small $\alpha$ local correlations will prevail
over the ergodic properties of the Ornstein--Uhlenbeck process. Of course by ``small'' we mean relative to
the value of $\lambda$.

In the very recent work \citet{LHoCAne2013} also consider a phylogenetic Ornstein--Uhlenbeck model.
They show that the maximum likelihood estimator of $\theta$ is not consistent. This does not
contradict our work as they differ from us on modelling assumptions. Firstly
they have a different model of tree growth, they 
assume a nested sequence of ultrametric trees with bounded internal node heights.
This means that tree $n-1$ is a subtree of tree $n$ and the tree height is bounded.
The second crucial difference is that the ancestral state
is not fixed but is a random variable drawn from the stationary distribution of 
the Ornstein--Uhlenbeck process. This combined with the boundedness of the tree height
implies that the correlation between any two species, regardless of the number of tips,
is bounded away from $0$ for all $n$ and hence the variance of the estimator of $\theta$ cannot
decrease to $0$.

In order to derive the above (using as in Paper III the laws of total variance and covariance)
we were required to compute the Laplace transforms of $U_{n}$ and $\tau^{(n)}$. 
Using them one can calculate all the moments of
$U_{n}$ and $(U_{n}-\tau^{(n)})$ and also relate our work
to the splitted nodal distance metric between phylogenies \citep{JFel2004,GCarMLlaFRosGVal2010}.

\subsection{Paper V --- Quantifying the effect of jumps at speciation}\label{sbsecPapV}
Papers III and IV assumed  Brownian motion and Ornstein--Uhlenbeck
evolution respectively. This is in line with the  idea that macroevolution is the consequence of microevolutionary
changes over many generations. However another possibility, indicated by fossil
records, has been pointed out, that phenotypes enjoy long periods
of stasis with rapid change concentrated in short time 
periods \citep{NEldSGou1972,SGouNEld1993}. The first type of evolution is
termed gradual, anagenetic or phyletic gradualism, the second cladogenetic or 
punctuated equilibrium. Naturally one can imagine that evolution could
be a combination of the two mechanisms and so a unified framework for them
should be developed. Such a framework has been introduced and developed
by \citet{FBok2002,FBok2008,FBok2010}. 
\citet{FBok2002} assumed that the logarithm of the phenotype evolves as a Brownian motion and at each
speciation event receives a normally distributed jump. 
Earlier \citet{AMooSVamDSch1999} used  both anagenetic and cladogenetic models in their Gruinae analysis.
Very recently \citet{JEasDWegCLeuLHar2013} model \textit{Anolis} lizards evolution by a Brownian motion with jumps 
inside branches.

Paper V combines
the tree--free approach of Papers III and IV with speciation jumps. 
We consider both the Brownian motion and Ornstein--Uhlenbeck models. In fact we 
show, as expected, that the Brownian motion results are limits of the Ornstein--Uhlenbeck
ones when $\alpha\to 0$. We calculate the interspecies correlation coefficient and
in addition we derive the probability generating functions of
the tree height, $U_{n}$ and time to coalescent of a random pair of tip species, $\tau^{(n)}$.
One can see in the Brownian motion case that increasing the variance of the jumps
decorrelates species while a numerical treatment of the correlation function
indicates the same in the Ornstein--Uhlenbeck case. 

One of the biological motivations for considering such models is to determine which
sort of evolution is dominating for a trait. As \citet[][]{AMooSVamDSch1999} point out speciational
change should be more relevant for  mate choice traits while anagenetic for traits under
continuous selection pressures often changing direction. \citet{FBok2002,FBok2008,FBok2010}
proposes different methods to estimate parameters of the models, however to quantify
effects of each type of evolution and to compare them we need to also know for how
long each type of evolution has occurred, e.g. if speciation was extremely rare
in some very old lineage then jumps might have had little opportunity to contribute.
Therefore in Paper V we suggest that the expectation (as we don't know the tree height
nor precise number of jumps in our framework) of the quadratic variation would be a good quantity.
In fact we show that it can be decomposed into the sum of the anagenetic and the cladogenetic
component and so we can calculate the (expected) proportion of change due to each mode of evolution.
In Paper V we illustrate this by interpreting the Hominoidea related estimates of  \citet{FBok2002,FBokVBriTSta2012}.

One of the requirements of this is that we know the number (or in our case expectation) of jumps occurring
along the (random) lineage. We are able to calculate this for the pure--birth tree 
\citep[see also][for the same formulae]{MSteAMcK2001} but had to resort to simulation methods
when a death component was included.  

Our results hold for a number of different jump models. The jump can have any distribution
provided it is mean zero and has a finite variance. It can occur after speciation on both lineages,
on a randomly selected one or on each one with a given probability.

L\'evy processes such as the Laplace motion \citep{KBar2012,MLanJSchMLia2013} are another
approach to the inclusion of evolutionary jumps however they are not as easy to interpret
as ``a jump at/due to/causing speciation''.

\subsection{Example application}\label{sbsecEx}
\begin{figure}[!h]
\centering
\includegraphics[width=5.4cm]{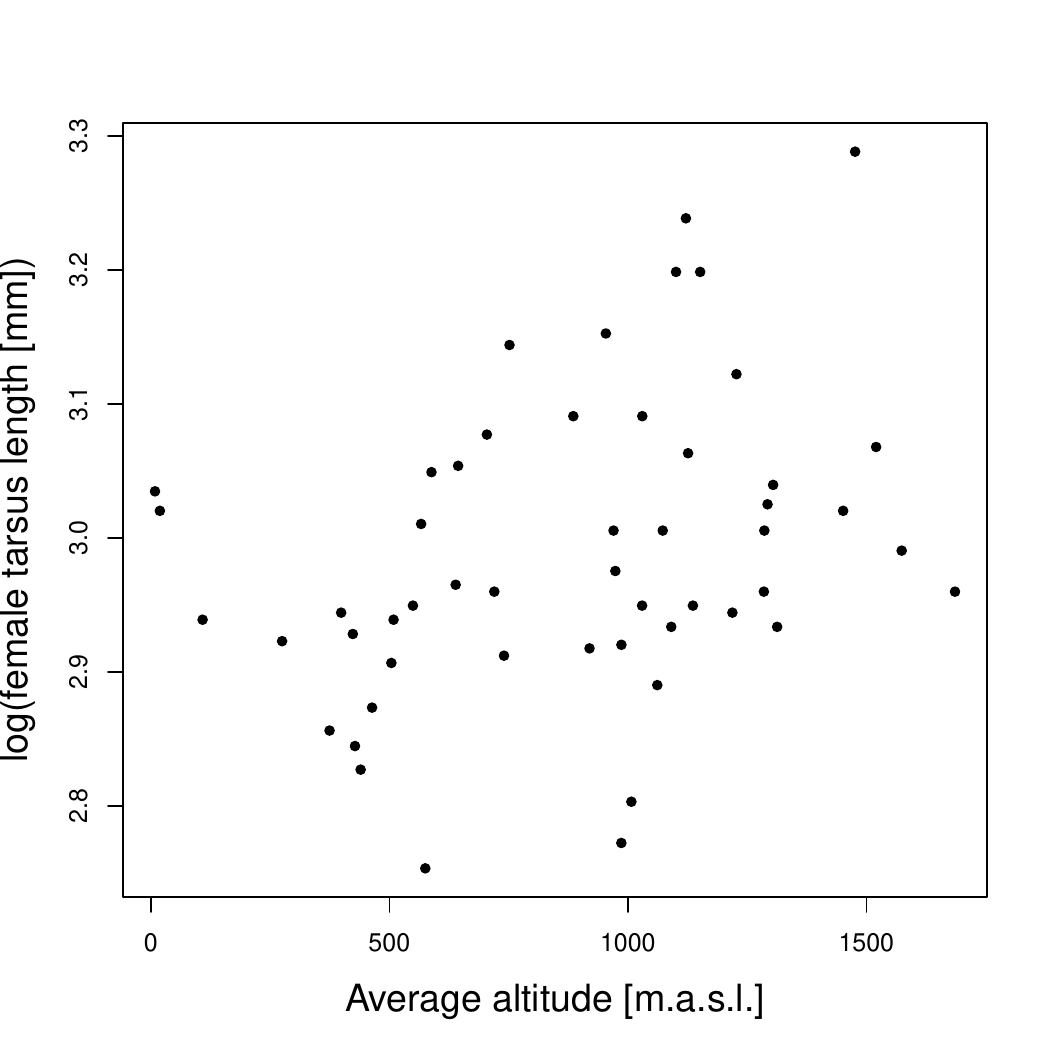}
\includegraphics[width=5.4cm]{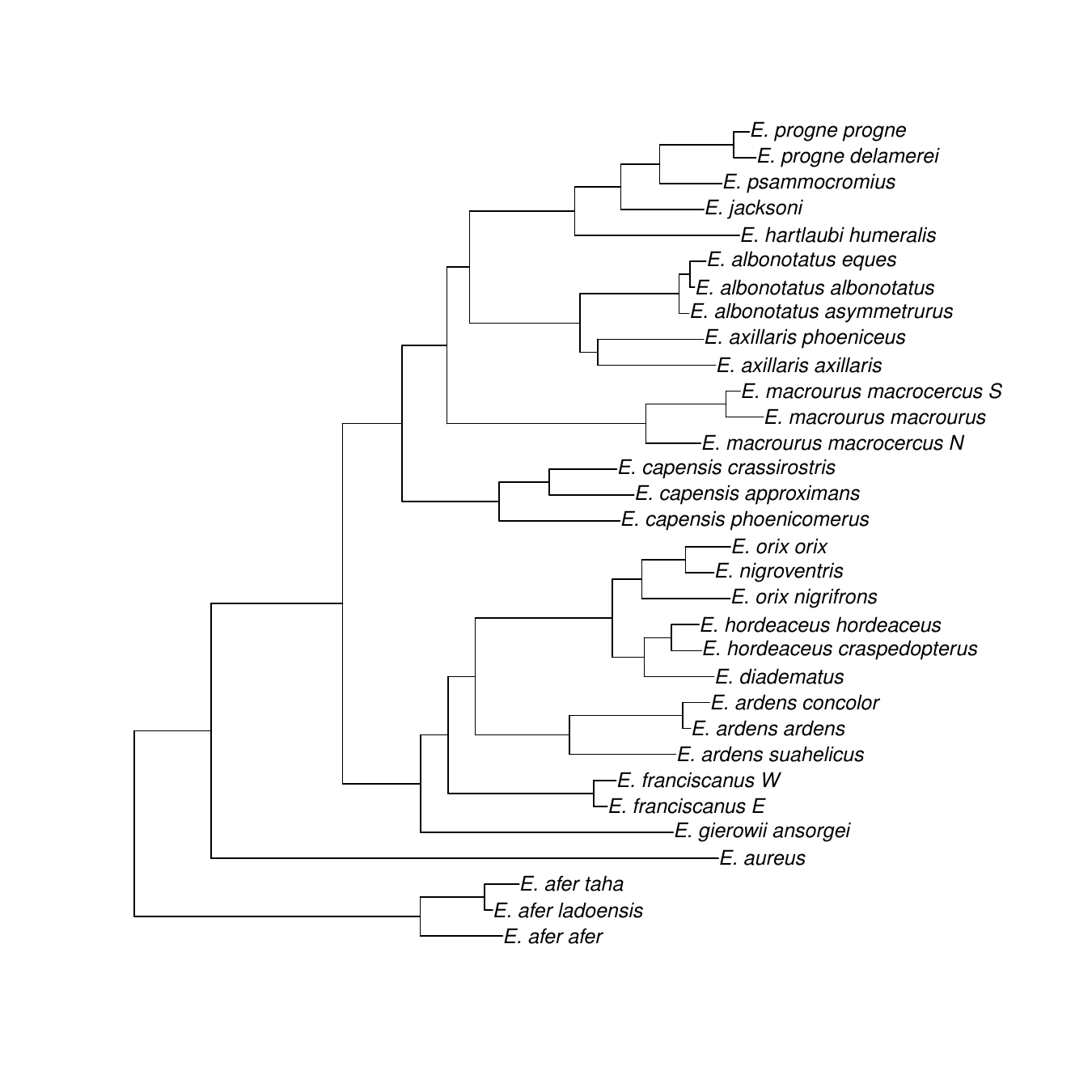}
\caption{
Left: logarithm of female tarsus length of the \textit{Ploceus} genus plotted as a function of average altitude.
Right: \textit{Euplectes} phylogeny.
}
\label{figPlEp}
\end{figure}

To illustrate and motivate the usefulness  of the tree--free approach 
we will conduct an example analysis of the logarithm of \textit{Ploceus} female tarsus length
(Fig. \ref{figPlEp} --- measurements provided by Staffan Andersson, 
Department of Biological and Environmental Sciences, University of Gothenburg, private communication). The
phylogeny of this genus is unknown so a usual comparative approach is not possible. 
We assume that the logarithm of the female tarsus evolves according to a single optimum 
Ornstein--Uhlenbeck process,
\bd
\ud X(t) = -\alpha (X(t) - \theta) \ud t + \sigma \ud W(t)
\ed
and would like to make inference on the value of $\theta$ from species average measurements of 
$52$ members of the \textit{Ploceus} genus. We showed in Paper IV that the 
sample average is a consistent and asymptotically unbiased estimator of $\theta$ and from
our data we obtain $\hat{\theta} = \overline{X}_{52} =  19.993$. We would then like to calculate
the phylogenetic confidence interval presented in Paper IV. There is a problem as since we do not
have the phylogeny, we cannot obtain an estimate of $\alpha$ and $\lambda$. However a $33$ species phylogeny 
of a sister clade, the \textit{Euplectes} genus
(both are members of the Ploceidae family),
is available \citep[Fig. \ref{figPlEp},][]{MPraEJohSAnd2008,MPraSAnd2009}
and we posses measurements of female tarsus length in $32$ species in this genus (measurements, including intra--species
variation, provided by Staffan Andersson, $\overline{X}_{32}=20.991$, $S^{2}_{32}=9.5$). 
Using the R laser package (function pureBirth) we estimate 
\mbox{$
\hat{\lambda}=6.146$} and using the mvSLOUCH package (Paper II)
we obtain that in this genus (assuming the single optimum Ornstein--Uhlenbeck model)
\mbox{$
\hat{\alpha}=8.719$,} 
\mbox{$
\widehat{\sigma^{2}}=166.248$,} 
\mbox{
$\hat{\theta}=21.158 \pm 1.646$} (regression based confidence intervals conditional on $\alpha$ and $\sigma$). 
As a check the 
stationary variance equals $\widehat{\sigma^{2}}/(2\hat{\alpha}) \approx 9.534$ 
(very close to $S^{2}_{32}$ despite the small sample size,
measurement error and that the phylogeny is not ultrametric).
We assume that these estimates are good enough to plug into the \textit{Ploceus} analysis
(however the sample variance in the \textit{Ploceus} genus is $S^{2}_{52}= 5.328$ so the results have
to be considered approximate). 
In Paper IV the phylogenetic confidence intervals were derived for $\lambda=1$
so in our formula we have to take instead of $\alpha$ 
the value $\alpha^{\ast} = \alpha/ \lambda$ 
estimated in our case as, $\hat{\alpha}/\hat{\lambda} \approx 1.419 > 0.5$. This means that we
are in the fast adaptation regime and the $95\%$  phylogenetic confidence interval will be given by the formula,
\bd
\overline{X}_{52} \pm \left(\sqrt{\frac{2\alpha^{\ast}+1}{2\alpha^{\ast}-1}} \right) \left(1.96 \frac{\sqrt{S^{2}_{52}}}{\sqrt{n}}\right)
\approx 19.993 \pm 0.91 .
\ed
The phylogenetic correction factor is $\sqrt{\frac{2\alpha^{\ast}+1}{2\alpha^{\ast}-1}}\approx 1.445$ 
so we can see that the phylogeny plays a role. The confidence interval
is rather tight so we can say that $19.993$ is a fair estimate of $\theta$. We can also see that the 
confidence intervals in the two genera overlap so the female tarsus length might be under
similar pressures in both.

Assuming that $S^{2}_{52}$ estimates the stationary variance well enough and taking $\hat{\alpha}=8.719$ we
obtain that \mbox{
$\widehat{\sigma^{2}}=2\hat{\alpha}S^{2}_{52}\approx 92.91$} in the \textit{Ploceus} genus. This means that
the main difference between the two genera (with respect to female tarsus length) 
is in the magnitude of the stochastic perturbations
to the adaption towards the primary optimum. They are greater in the \textit{Euplectes} and so we should expect
more diversity in this clade. 

The above analysis illustrates our tree--free method: in what situation it can be useful and
where more work needs to be done. We can suspect that in the \textit{Ploceus} the optimal value for the 
female tarsus length should be dependent on some environmental variables, e.g. Fig. \ref{figPlEp} suggests the altitude
of the species' habitat. The current method cannot take this into account so this indicates a direction 
of future development. The phylogenetic confidence interval depends on knowing $\alpha$ --- we cannot 
estimate it as the phylogeny of our clade of interest is unknown. Therefore we have to obtain it
by other means --- here we estimated it from a sister clade with the hope that it would not differ 
substantially from the one in our clade of interest. It would therefore be desirable to incorporate
uncertainty in the estimation of $\alpha$ into the method.

\clearpage
\newpage
\section{Phylogenetic networks}
\subsection{Introducing networks}
Evolutionary relationships between species are traditionally portrayed in a tree like structure.
We are however more and more aware that this is not the whole story. There are 
cases, especially with many plant species 
e.g. most breeds of wheat, peppermint (hybrid between spearmint and watermint),
grapefruit (hybrid between pomelo and Jamaican sweet orange),
where a more realistic description is a phylogenetic network due to hybridization events.
In terms of phylogenetic comparative methods considering networks instead of trees can be useful 
in a situation where we are studying a phenotype evolving in separated populations
(allowed however to mix from time to time) that could be under different environmental
pressures. 

The area of phylogenetic networks is a new topic.
There is a large number of different types of networks. As in the case of trees
one can consider rooted or unrooted networks. The fundamental concept related to
unrooted networks is that of splits. A split is a bipartition of the set of taxa
such that the two sets are disjoint. Combined they are the set of taxa. In  
a phylogenetic tree every edge uniquely corresponds to a split.  
In the network case every split is represented by a collection of edges.
Rooted phylogenetic networks can be used to represent hybridization, 
recombination and reassortment events. They can be also used to reconcile species
and gene trees if duplication, loss or transfer events took place. More
generally they can be used to represent a set of competing
trees on the same set or overlapping sets of taxa.
\citet{DHusRRupCSco2010} in their very recent
work discuss the different network types and the algorithms presently being used for 
their reconstruction. They concentrate on non--probabilistic network models
and reconstruction methods. Another approach is of course
a probabilistic one. \citet{GJonSSagBOxe2013} develop a Bayesian methodology
for hybridization network reconstruction. In such a case  a 
prior distribution for the time to hybridization would be desirable
and Paper VI is an attempt at this.

\subsection{Paper VI --- Time to hybridization is approximately exponential}\label{sbsecPapVI}
In Paper VI we assume a very simple network setup. 
Firstly we assume that there are $n$ contemporary diploid species linked by a pure--birth tree
with speciation rate $\lambda$. Then we model hybridization by a Poisson process. If during a 
time period $t$ there are $k\ge 2$ diploid species then the number of hybridization events, $N_{k}(t)$,
in this time period is Poisson distributed (parametrized by $\beta >0$),
\be
\prob{N_{k}(t)}{j} = \frac{\left(\binom{k}{2}\beta t\right)^{j}}{j!}e^{-\binom{k}{2}\beta t}, ~~~ j=0,1,2,\ldots~.
\ee
We condition on there being exactly one hybridization event inside the tree induced by the 
$n$ diploid species. We make no statements about what occurs on the lineage(s) started from this
hybridization event. There are two parameters in this model the speciation rate $\lambda$
and hybridization rate $\beta$, however the distribution of the random variable
$\tau_{n}$ --- the time to the singled--out hybridization event can be
characterized in terms of $\lambda$ and the compound parameter (a relative hybridization rate),
\be
\gamma := \frac{\beta}{2\lambda}.
\ee
We define the random variable $\kappa_{n} \in \{2,\dots,n \}$ the number of species present when the 
hybridization event occurred and then decompose,
\be
\tau_{n} = X + \sum\limits_{j=\kappa_{n}+1}^{n} T_{j},
\ee
where $T_{i}$ is the random time during which there were exactly $i$ species present,
distributed exponentially with rate $i\lambda$ as the minimum of $i$ independent
exponential random variables with rate $\lambda$, see Fig. 1 in Paper VI and $Y$ is uniformly distributed on the
interval $[0,T_{\kappa_{n}}]$.
Using this decomposition we obtain all the moments of $\tau_{n}$,
\be
\begin{array}{ll}
\E{\tau_{n}^{r}\vert \tau_{n} < \infty} = &
\frac{1}{\sum\limits_{k=1}^{n-1}\frac{k\gamma}{1+k\gamma}}\frac{r!}{\lambda^{r}}
\sum\limits_{k=1}^{n-1}\frac{k\gamma}{1+k\gamma}
\\ & \times
\sum\limits_{i_{1}=k}^{n-1}\sum\limits_{i_{2}=i_{1}}^{n-1}\ldots\sum\limits_{i_{r}=i_{r-1}}^{n-1}
\frac{1}{(1+i_{1})(1+\gamma i_{1})}\cdot\ldots\cdot\frac{1}{(1+i_{r})(1+\gamma i_{r})}.
\end{array}
\ee
In addition 
assuming that $\lambda$ is constant, $n\gamma\to 0$ and proving the identity,
\be
\sum\limits_{k=1}^{n-1}k\sum\limits_{i_{1}=k}^{n-1}\ldots\sum\limits_{i_{r}=i_{r-1}}^{n-1}\left(\frac{1}{1+i_{1}}\cdot \ldots \frac{1}{1+i_{r}} \right) = 2^{-r}\binom{n}{2},
\ee
we obtain that conditional on $ \tau_{n} < \infty$,
\be
\tau_{n} \stackrel{\mathcal{D}}{\xrightarrow{\hspace*{1cm}}} \mathrm{Exp}(2\lambda).
\ee
This weak convergence is due to for each $r$, 
\mbox{
$\E{\tau_{n}^{r}\vert \tau_{n} < \infty} \to \frac{r!}{(2\lambda)^{r}}$}, the $r$--th moment
of the exponential with $2\lambda$ rate distribution.
In Paper VI we also show via simulations (histograms in Fig. 4 therein) that the exponential
approximation is good for $n$ even as small as $5$ or $10$.

In practice we cannot expect to know the parameter $\gamma$ and it should be estimated
from the observed data --- $n$ in our case. The method of moments estimator is 
$\tilde{\gamma}_{n} = 1/\binom{n}{2}$ and for $n\ge 4$ we have the following bounds
on the maximum likelihood estimator, $\hat{\gamma}_{n}$,
\be
\tilde{\gamma}_{n}=\frac{2}{n(n-1)} \le \hat{\gamma}_{n} \le \frac{2}{n(n-3)}
\ee
with the likelihood function given by,
\be
\mathrm{P}_{\gamma}(\tau_{n} < \infty) =  \prod\limits_{i=1}^{n-1}\frac{1}{1+i\gamma} \sum\limits_{k=1}^{n-1}\frac{k\gamma}{1+k\gamma}.
\ee

\clearpage
\newpage

\section{Future developments}\label{secFuture}
\citet{KBarLic2011} discussed that one of the ways
of building on the presented there work
was to get around the need to condition on a pre--estimated phylogenetic tree, i.e.
consider a tree--free model.
Papers III, IV and V do exactly that. 
They can be of course further expanded by moving into a multivariate setting (with e.g. predictor/response traits) or
allowing for (in the Yule--Ornstein--Uhlenbeck setting) multiple adaptive peaks developing as a Markov chain.
Allowing for hypothesis testing concerning the number of adaptive peaks in this framework
would be very desirable from a biological point of view, as current  
comparative analysis methods essentially condition on this
and their layout on the tree. To work with a multitrait tree--free comparative 
model we would need to understand how the different traits influence each other
along the unobserved tree and what information on these interactions is contained
in the contemporary sample. A starting point is the study of the correlation coefficient
in a bivariate neutrally (Brownian motion) evolving trait. Other analysis specific 
extensions have been discussed in the example application, Section \ref{sbsecEx}.

One of the main modelling assumptions of the underlying framework of
conditioned branching processes is that we know the number of contemporary
species. While this seems reasonable at first, we are still in the process
of discovering  new (or reclassifying) species 
\citep[primarily in the lower orders (e.g. simple invertebrates, primitive plants) but there 
is a very recent example of a new antelope][]{MColetal2010}.
Therefore it might be desirable to consider that we have only observed
a certain fraction of the species and so can only have an induced sub--tree. 
\citet{TSta2009,TSta2011a,SHohTStaFRonTBri2011,ALamHAleTSta2013,ALamHMorREti2013} 
have already considered questions related to this in different settings.
The first step in combining incomplete species sampling with phylogenetic
comparative methods could be a study of the behaviour of
the interspecies correlation coefficient in this situation. 
The study of inter--coalescent times of a subtree begun in Paper IV could 
be a starting point for this.

In Paper V we studied the effects of jumps on the sample mean and variance estimators.
An interesting question is how do the jumps change the limit theorems of 
Paper IV and hence the form of the phylogenetic confidence intervals. 
The jump component in Paper V can have any distribution (provided it is mean $0$ 
and finite variance). One question is what further properties would be needed
for the central limit theorem to still hold in the case of fast adaptation with jumps.

Both Papers IV and V assume that there are no death events, this being naturally unrealistic.
The compact form of the interspecies correlation coefficient of Paper III gives hope that
further properties of a tree--free model with a death component can be derived
analytically and so this is a further research path.

Paper VI can be immediately expanded by including death events, allowing for a specific 
number of hybridization events/contemporary hybrid species. Combining its results with the
tree--free models would allow us to study phenotypes evolving on a
network structure instead of a tree.

One can find in Papers III and V 
links between tree properties and particular phylogenetic indices/metrics.
A natural continuation of this is a systematic approach to go through
these and other common summary functions associated with trees
and study how our results can be applied to them. 

The other directions indicated by \citet{KBarLic2011} still remain open.
The results of Paper I can be developed in the direction of studying the 
effects of measurement error on the estimation of covariance structures.

We can develop Paper II in a couple of directions.
The first one concerns modelling
trait evolution and interactions among the traits and the environment.
An initial step can be to relax the regimes allowing them to change at random times thus forming a Markov chain.
Secondly one can try to address the various
statistical questions related to phylogenetically 
structured observations.
Paper IV partially addresses this by deriving a phylogenetic
confidence interval for a constant on the whole tree optimal trait value.
Including the measurement error correction
procedures of Paper I into comparative model estimation programs
is also an interesting software development direction.

Both the traditional phylogenetic comparative approach
and our discussed here tree--free approach assume
that there is a one way flow of effects from speciation (phylogeny)
to the phenotype. This is of course biologically unrealistic as 
one would expect some phenotypic explanation for speciation.
Therefore the incorporation of a feedback mechanism from the trait process to the 
branching process in our tree--free framework is another future goal.

\clearpage
\newpage